\documentclass[%
 aip,
 reprint,
 superscriptaddress,
 amsmath,
 amssymb
]{revtex4-2}

\usepackage{listings}
\usepackage{graphicx}
\usepackage{enumitem}
\usepackage[T1]{fontenc}
\usepackage{bm}
\newcommand{\argmin}[1]{\underset{#1}{\operatorname{arg}\,\operatorname{min}}\;}
\usepackage[scale=7,opacity=0.2,color=gray]{background}
\backgroundsetup{contents={\sffamily Accepted Manuscript}}

\begin{document}

\title{A single-shot measurement of time-dependent diffusion over sub-millisecond timescales using static field gradient NMR}

\author{Teddy X. Cai}
\email{teddy.cai@nih.gov}
\affiliation{Section on Quantitative Imaging and Tissue Sciences, Eunice Kennedy Shriver National Institute of Child Health and Human Development, National Institutes of Health, Bethesda, Maryland 20892, USA}
\affiliation{Wellcome Centre for Integrative Neuroimaging, Nuffield Department of Clinical Neurosciences, University of Oxford, Oxford, UK}
\author{Nathan H. Williamson}
\affiliation{Section on Quantitative Imaging and Tissue Sciences, Eunice Kennedy Shriver National Institute of Child Health and Human Development, National Institutes of Health, Bethesda, Maryland 20892, USA}
\affiliation{National Institute of General Medical Sciences, National Institutes of Health, Bethesda, Maryland 20892, USA}
\author{Velencia J. Witherspoon}
\affiliation{Section on Quantitative Imaging and Tissue Sciences, Eunice Kennedy Shriver National Institute of Child Health and Human Development, National Institutes of Health, Bethesda, Maryland 20892, USA}
\author{Rea Ravin}
\affiliation{Section on Quantitative Imaging and Tissue Sciences, Eunice Kennedy Shriver National Institute of Child Health and Human Development, National Institutes of Health, Bethesda, Maryland 20892, USA}
\affiliation{Celoptics, Rockville, Maryland 20852, USA}
\author{Peter J. Basser}
\email{peter.basser@nih.gov}
\affiliation{Section on Quantitative Imaging and Tissue Sciences, Eunice Kennedy Shriver National Institute of Child Health and Human Development, National Institutes of Health, Bethesda, Maryland 20892, USA}

\date{\today}

\begin{abstract}
Time-dependent diffusion behavior is probed over sub-millisecond timescales in a single shot using an NMR static gradient, time-incremented echo train acquisition (SG-TIETA) framework. The method extends the Carr-Purcell-Meiboom-Gill (CPMG) cycle under a static field gradient by discretely incrementing the $\pi$-pulse spacings to simultaneously avoid off-resonance effects and probe a range of timescales ($50-500 \;\mu$s). Pulse spacings are optimized based on a derived ruleset. The remaining effects of pulse inaccuracy are examined and found to be consistent across pure liquids of different diffusivities: water, decane, and octanol-1. A pulse accuracy correction is developed. Instantaneous diffusivity, $D_{\mathrm{inst}}(t)$, curves (i.e., half of the time derivative of the mean-squared displacement in the gradient direction), are recovered from pulse accuracy-corrected SG-TIETA decays using a model-free, log-linear least squares inversion method validated by Monte Carlo simulations. A signal-averaged, 1-minute experiment is described. A flat $D_{\mathrm{inst}}(t)$ is measured on pure dodecamethylcyclohexasiloxane whereas decreasing $D_{\mathrm{inst}}(t)$ are measured on yeast suspensions, consistent with the expected short-time $D_{\mathrm{inst}}(t)$ behavior for confining microstructural barriers on the order of microns.
\end{abstract}
\maketitle 

\section{Introduction}
As molecules diffuse, they interact with their surroundings and ``[feel] the boundary'' \cite{Kac1966}, causing their ensemble displacement behavior to be influenced by the morphology of the microenvironment. More specifically, long-range correlations such as confining barriers impart a nonlinear time-dependence to the ensemble-averaged net mean-squared displacement, $\langle \textbf{r}^2(t) \rangle$ (in $\mathbb{R}^3$). This leads to a time-dependent diffusion coefficient \cite{Mitra1992, Mitra1993},
\begin{equation}\label{eq: D(t) def}
    D(t) =\frac{\langle \textbf{r}^2(t) \rangle}{6t} \equiv \frac{1}{3t}\int_0^{t} (t - t') \,\mathrm{tr} \bm{(} \boldsymbol{\mathcal{D}}(t')\bm{)}\, dt',
\end{equation}
where $\boldsymbol{\mathcal{D}}(t') = H(t')\langle \textbf{v}(t') \textbf{v}^{\mathrm{T}}(0) \rangle \equiv \partial_t^2 \left[H(t) \langle \textbf{r}(t) \textbf{r}^{\mathrm{T}}(0) \rangle\right]$ is the causal velocity autocorrelation tensor, $H(t)$ is the unit step function, and ``tr'' is the trace operation. 

Microstructural features can be inferred from the behavior of $D(t)$ at limiting short and long timescales (see Sen \cite{Sen2004} and Reynaud \cite{Reynaud2017} for review). At the short-time limit, $D(t)$ exhibits universal behavior \cite{Mitra1992} which depends on the barrier surface-to-volume ratio, $S/V$,
\begin{equation}\label{eq: Mitra short-time}
    D(t) \simeq D_0\left[1 - \frac{S}{V}\left(\frac{4 \ell_D}{9\sqrt{\pi}}\right) \right], \; t\rightarrow 0,
\end{equation}
where $\ell_D = \sqrt{D_0t}$ is the diffusion length scale and $D_0 \equiv D(t)|_{t = 0}$ is the free diffusivity. As $\ell_D$ increases, the barrier permeability, $\kappa$, may begin to affect $D(t)$. \cite{Tanner1978, Tanner1979, Sen2003, Sen2004b} While $\ell_D$ remains short, barriers appear flat to the small fraction of nearby walkers that encounter them \cite{Sen2003, Novikov2011b}, introducing a linear $\kappa t$ term in Eq. \eqref{eq: Mitra short-time},
\begin{equation}\label{eq: Sen permeability short-time}
    D(t) \simeq D_0\left[1 - \frac{S}{V}\left(\frac{4 \ell_D}{9\sqrt{\pi}} - \kappa t\right) \right], \; t \ll \tau_D,
\end{equation}
where $\tau_D = \bar{a}^2/(2D_0)$ is the time to diffuse across the mean pore of size $\bar{a} = 6V/S$. Curvature \cite{Sen2003} and surface-relaxivity \cite{Mitra1993} may also affect $D(t)$. Tortuosity principally affects the long-time $D(t)$ \cite{Mair2002, Latour1993} and can be categorized into disorder classes with structural exponent, $p$. \cite{Novikov2011b, Novikov2014} The long-time $D(t)$ follows a $p$-dependent power law \cite{Novikov2014},
\begin{equation}\label{eq: long-time power law}
    D(t) \simeq D_\infty+\mathrm{const.}\,\cdot t^{-\vartheta}, \; t\rightarrow\infty,
\end{equation}
where $D_\infty =\lim_{t\rightarrow \infty} D(t)$ and $\vartheta = (p + 3)/2$. 

Diffusion-weighted (DW) nuclear magnetic resonance (NMR) methods are highly sensitive to $\langle \textbf{r}^2(t) \rangle$ \cite{Hahn1950, Woessner1963, Stejskal1965, Neuman1974, Callaghan1991, Price2009, Keeler2010}, and provide a powerful means to probe rich $D(t)$ behaviors and infer distinct microsctructural features. DW-NMR experiments have been used to study the short- and long-time $D(t)$ in porous media ranging from sedimentary rock to skeletal muscle. \cite{Hurlimann1996, Latour1993, Latour1994, Callaghan1995, Schacter2000, Sigmund2014, Novikov2014, Reynaud2017}

The NMR spin echo dephasing, however, is not simply written in terms of $D(t)$ itself. Instead, the echo dephasing is often expressed in terms of the real part, $\Re$, of the Fourier transform of $\boldsymbol{\mathcal{D}}(t)$. \cite{Seymour1997, Khrapitchev2003, Novikov2011} From Eq. \eqref{eq: D(t) def} \cite{Novikov2011},
\begin{equation}\label{eq: D(omega) def}
    \frac{\mathrm{tr}\bm{(}\Re\left[\boldsymbol{\mathcal{D}}(\omega)\right]\bm{)}}{3} = D_0 + \int_0^{\infty}  \partial_t^2 \left[ tD(t) \right] e^{i\omega t} dt.
\end{equation}
High and low frequency $\mathrm{tr}\bm{(}\Re\left[\boldsymbol{\mathcal{D}}(\omega)\right]\bm{)}$ behaviors reveal the short- and long-time $D(t)$, respectively. The relationship between $\Re\left[\boldsymbol{\mathcal{D}}(\omega)\right]$ and the ensemble echo dephasing follows from a cumulant expansion of the phase expectation value and a Gaussian phase distribution approximation \cite{Douglass1958}, yielding the normalized echo intensity \cite{Stepisnik1981, Stepisnik1993}, 
\begin{equation} \label{eq: MGSE}
    \frac{I(T)}{I_0} = \exp\left(-\frac{1}{\pi}\int_0^{\infty}  \textbf{F}^{\mathrm{T}}(\omega)\, \Re\left[\boldsymbol{\mathcal{D}}(\omega)\right] \textbf{F}(\omega) \,d\omega\right), 
\end{equation}
where $\textbf{F}(\omega)$ is the truncated Fourier transform of $\textbf{F}(t)$ at the echo time, $T$, (i.e., $\textbf{F}(T) = \textbf{0}$),
\begin{equation}\label{eq: F(omega def)}
    \textbf{F}(\omega) = \int_0^T  \textbf{F}(t)  e^{i\omega t} dt, 
\end{equation}
$\textbf{F}(t) = \gamma \int_0^t \textbf{G}(t') dt'$ , $\textbf{G}(t)$ is the gradient waveform, and $\gamma$ is the gyromagnetic ratio. The assumed Gaussian phase approximation is valid for most relevant experimental cases (cf. Stepi\v{s}nik \cite{Stepisnik1999}). \cite{Axelrod2001, Sukstanskii2003}

Eqs. \eqref{eq: MGSE} and \eqref{eq: F(omega def)} show that spectral tuning of $|\textbf{F}(\omega)|^2$ results in narrow sampling of $\Re\left[\boldsymbol{\mathcal{D}}(\omega)\right]$ in the gradient direction, $\hat{\textbf{g}}$. $\textbf{G}(t)$ can be periodically time-modulated \cite{Callaghan1996} (e.g., by using a sinusoidal $\textbf{G}(t)$ \cite{Schacter2000}) so that the spectral density of $|\textbf{F}(\omega)|^2$ concentrates near some frequency, $\omega_F$. In this case, $I(T)/I_0$ becomes well-approximated by $\exp{\bm{(}-b(T)\times \hat{\textbf{g}}^{\mathrm{T}} \Re\left[\boldsymbol{\mathcal{D}}(\omega_F)\right]\hat{\textbf{g}}\bm{)}}$, where \cite{Stejskal1965}
\begin{equation}
b(T) = \int_{0}^{T}\left| \textbf{F}(t) \right|^2 dt \equiv \frac{1}{\pi}\int_0^{\infty} |\textbf{F}(\omega, T)|^2  d\omega. 
\end{equation}
Individual experiments become a point-wise sampling of $\hat{\textbf{g}}^{\mathrm{T}}\Re\left[\boldsymbol{\mathcal{D}}(\omega_F)\right]\hat{\textbf{g}}$. This ``temporal diffusion spectroscopy'' \cite{Gore2010} approach is robust, but has limited time resolution because it individually probes $\omega_F$. Furthermore, the shortest probe-able timescale (i.e., largest $\omega_F$) is limited to about 10 ms by the pulsed gradient hardware.\cite{Reynaud2017, Gore2010} An alternative approach is needed for the real-time study of $D(t)$ \emph{across} timescales and to reach the information-rich, short-time regime ($\lesssim 1$ ms) in biological systems. Static gradient (SG) hardware permits extremely fast cycling of the effective gradient direction using radiofrequency (RF) $\pi$-pulse trains and thereby provides access to these timescales.\cite{Callaghan1995, Callaghan1996} Here, we extend the classical Carr-Purcell-Meiboom-Gill (CPMG)\cite{Carr1954, Meiboom1958} experiment under an SG in the RF field to probe the time-varying, sub-millisecond diffusivity in one shot.  

\section{Theory}
\subsection{Time-dependent signal representation}

\begin{figure}
\includegraphics{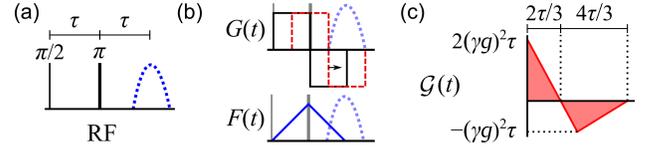}
\caption{\label{fig: gradient corr calc} Example $\mathcal{G}(t)$ calculation. (a) Radiofrequency (RF) pulses in a static gradient of amplitude $g$. (b) $G(t)$, shifted $G(t+s)$ (red, dashed), and $F(t)$. (c) Corresponding $\mathcal{G}(t)$. $G(t)$ is assumed to be zero outside of $t = [0, 2\tau]$.}
\end{figure} 

To begin, an alternative signal representation is used. The echo attenuation is related to the stationary position autocorrelation tensor, $\boldsymbol{\mathcal{R}}(t, t') = \langle \textbf{r}(t) \textbf{r}^{\mathrm{T}}(t') \rangle \equiv \boldsymbol{\mathcal{R}}(|t - t'|)$ -- again assuming a Gaussian phase distribution \cite{Stepisnik1981, Stepisnik1993, Ning2017} -- by
\begin{equation} \label{eq: position autocorr and G 1-D}
    \frac{I(T)}{I_0} = \exp\left(-\frac{\gamma^2}{2}\int_0^{T}\int_0^{T} \textbf{G}^{\mathrm{T}}(t)\boldsymbol{\mathcal{R}}(t, t')\textbf{G}(t') \, dt dt'\right),
\end{equation}
Eq. \eqref{eq: position autocorr and G 1-D} can be rewritten according to Ning et al. \cite{Ning2017} by integrating along the level set of $t - t'$. For unidirectional encoding, i.e., $G(t) = ||\textbf{G}(t)||$ and $F(t) = \gamma\int_0^t G(t')\, dt'$,
\begin{equation}\label{eq: C(t) and D_inst}
    \frac{I(T)}{I_0} = \exp\left(-\int_0^T \mathcal{C}(t) D_\mathrm{inst}(t)\, dt\right),
\end{equation}
where $D_\mathrm{inst}(t)$ is the \emph{instantaneous} diffusivity along the gradient direction $\hat{\textbf{g}}$,
\begin{equation}\label{eq: D_inst def}
D_\mathrm{inst}(t) := \partial_t \left[\frac{\hat{\textbf{g}}^{\mathrm{T}}\boldsymbol{\mathcal{R}}(t)\,\hat{\textbf{g}}}{2}\right] \equiv \partial_t \left[\frac{\langle \left[\textbf{r}(t)\cdot \hat{\textbf{g}}\,\right]^2 \rangle}{2}\right], \; t > 0,
\end{equation}
(dropping $\hat{\textbf{g}}$ as implied) and $\mathcal{C}(t)$ is the cumulative integral of the $\gamma G(t)$ autocorrelation function,  
\begin{equation}\label{eq: G(t) def}
\begin{gathered}
    \mathcal{C}(t) = \int_0^t \mathcal{G}(t')dt' \\
    \mathcal{G}(t') =\gamma^2\int_0^{T}G(t')G(t'+s)\,ds
\end{gathered},
\end{equation}
schematized in Fig. \ref{fig: gradient corr calc}. The attenuation becomes a sampling of $D_\mathrm{inst}(t)$ weighted by $\mathcal{C}(t)$, similar to how Eq. \eqref{eq: MGSE} describes a sampling of $\Re\left[\boldsymbol{\mathcal{D}}(\omega)\right]$ by $|\textbf{F}(\omega)|^2$.

\subsection{Recasting the problem}

Eq. \eqref{eq: C(t) and D_inst} is advantageous compared to Eq. \eqref{eq: MGSE} because $\mathcal{C}(t)$ is simple for general, non-periodic $G(t)$. The ill-posed inverse problem of finding $D_\mathrm{inst}(t)$ from many trivial $G(t)$ is tractable. For a train of $N$ echoes refocusing at times $T_n$, $\mathcal{C}_n(t)$ can be evaluated for each inter-echo attenuation $I(T_n)/I(T_{n-1})$ ($T_0 = 0$); i.e., each pair of adjacent echoes can be treated as an independent spin echo diffusion measurement. The problem is then recast as a weighted and regularized log-linear least squares (LLS) inversion by discretizing the time domain into $K$ bins of variable width, $\Delta t(k)$:
\begin{equation}\label{eqn: lin prog}
    \argmin{\textbf{X}} \; || \textbf{W}^{1/2}\left(\textbf{A}\textbf{X} - \textbf{B}\right)||_2^2 + \lambda||\boldsymbol{\Gamma} \textbf{X} ||_2^2,
\end{equation}
with coefficients
\begin{equation}\nonumber
    \textbf{A} = \begin{bmatrix}
    \int_{0}^{\Delta t(1)} \mathcal{C}_1(t)\, dt & \hdots & \int_{\Delta t(K-1)}^{\Delta t(K)} \mathcal{C}_1(t) \,dt \\ 
    \vdots &  & \vdots \\ 
    \int_{0}^{\Delta t(1)} \mathcal{C}_N(t)\, dt & \hdots & \int_{\Delta t(K-1)}^{\Delta t(K)} \mathcal{C}_N(t) \,dt \\ 
    \end{bmatrix},
\end{equation}
where \textbf{X} consists of time-interval $D_\mathrm{inst}(t)$ averages,
\begin{equation} \nonumber
    \textbf{X} = \begin{bmatrix}
    \frac{1}{\Delta t(1)}\int_{0}^{\Delta t(1)}D_{\mathrm{inst}}(t)\, dt \\ 
    \vdots  \\ 
    \frac{1}{\Delta t(K)-\Delta t(K-1)} \int_{\Delta t(K-1)}^{\Delta t(K)} D_{\mathrm{inst}}(t)\, dt
    \end{bmatrix}, 
\end{equation}
and $\textbf{B}^{\mathrm{T}} = -\ln\,\begin{bmatrix} I(T_1)/I_0 & \hdots & I(T_N)/I(T_{N-1}) \end{bmatrix}.$ The regularization matrix, $\boldsymbol{\Gamma}$, is chosen to consist of first and second-order finite difference matrices, reflecting an \textit{a priori} assumption of the smoothness and concavity of $D_{\mathrm{inst}}(t)$. The choice of $N$ is dictated by when the echo signal decays to the noise floor. The choice of $K$ and $\Delta t(k)$ is more arbitrary. As a preliminary heuristic, $K$ should be similar in magnitude to $N$, and $\Delta t(k)$ should be chosen such that (1) $D_\mathrm{inst}(t)$ does not vary greatly over any interval and (2) the $\mathcal{C}_n(t)$ integrals that comprise the entries of \textbf{A} are appreciable. Considering the behavior of Eqs. \eqref{eq: Sen permeability short-time} and \eqref{eq: long-time power law}, $\Delta t(k)$ should start out small and may gradually lengthen. The norm is weighted by a proportionality of the signal-to-noise ratio (SNR); since \textbf{B} consists of log ratios, the appropriate weights matrix, \textbf{W}, is an $N \times N$ matrix of signal differences, i.e., $I(T_{n-1}) - I(T_n)$. In this way, $D_\mathrm{inst}(t)$ can be estimated from a single echo train with varied $\mathcal{C}_n(t)$.

The motivating question of this Communication is as follows: Can a DW-NMR method probe the time-varying diffusivity in real-time? With Eq. \eqref{eqn: lin prog} in mind, the question can be separated into two parts: (1) How can $\mathcal{C}_n(t)$ of various time sensitivities be produced in one echo train? (2) How can every echo be made accurate? The answer to the first part follows from a well-known DW-NMR protocol. As mentioned, the SG-CPMG experiment can produce rapid effective gradient oscillation characterized by a triangle wave $F(t)$ with $\omega_F = (\pi/2\tau$) rad/s, where $2\tau$ is the spacing between $\pi$-pulses. This SG-CPMG $\omega_F$ (up to tens of kHz \cite{Callaghan1995}) exceeds that which is attainable with oscillating or pulsed field gradient (PFG) methods (up to $\sim100$ Hz). As a result, the SG-CPMG method is uniquely able to probe the short-time diffusion regime in small ($\bar{a} \lesssim \mu$m) structures. \cite{Callaghan1995, Callaghan1996, Zielinski2005, Lasic2006, Stepisnik2006, Zielinski2003, Song2003} Stimulated echoes represent another candidate DW-NMR method, but are ill-suited to single-shot, multi-echo acquisitions due to the signal loss inherent to $\pi/2$-pulses. 

\begin{figure}
\includegraphics{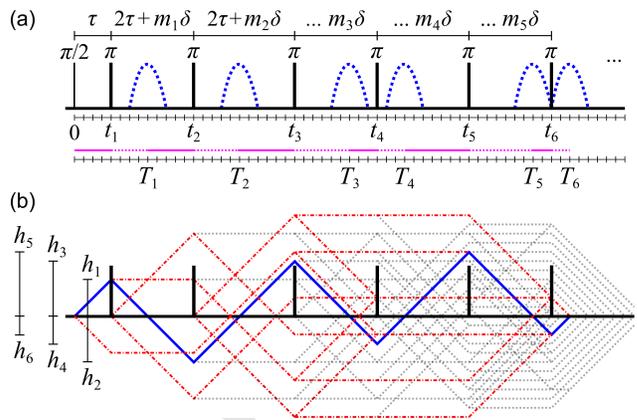}
\caption{\label{fig: pulse seq} Example SG-TIETA sequence. (a) Timings: $m_j = \{1, 3, 1, 2, 1\}$, $\tau = 4\delta$, and $\delta = 14\; \mu$s $= 1$ dash. $\pi$-pulses occur at $t_n$ and direct echoes (blue lines) form at $T_n$. Magenta line indicates timing behavior: $T_n = t_{n} + h_n$, where $h_n$ is the normalized $|F(t)|/\gamma g$ ``height'' at $t_n$, given by $h_1 = \tau$ and $h_{n} = 2\tau + m_n\delta -h_{n-1}$ for $n > 1$. (b) Direct echo $F(t)$ and other coherence pathways that refocus (red, dash-dot) or do not refocus (gray, dotted). Relative values of $h_n$ are indicated.}
\end{figure}

Varying the spacing of the $\pi$-pulses in an SG-CPMG styled acquisition is thus the preferred method to produce various $\mathcal{C}_n(t)$ in one shot. Others have explored the concept of modifying SG-CPMG pulse spacings to measure time-varying diffusion, but ultimately retained a $\pi$-pulse train with repeated spacing.\cite{Song2003} We extend such methods by modifying \emph{every} $\pi$-pulse spacing. Spacings can be incremented to retain signal. We choose the spacing between $\pi$-pulses to take the form: $2\tau + m_j \delta$, where $j$ indexes the $\pi$-pulse-to-pulse spacing, $m_j \in \mathbb{N}$, and $\delta$ is a unit time increment. We term this discrete spacing method the SG, time-incremented echo train acquisition (SG-TIETA), e.g., Fig. \ref{fig: pulse seq}. For SG-TIETA,
\begin{equation}\label{eq: C(t) SG-TIETA}
\mathcal{C}_n(t)= \gamma^2g^2 
\begin{cases}
 t \left( - \frac{3}{2}t + 2h_n \right) &  0 \leq t \leq h_n \\ 
 t \left(\frac{1}{2}t -h_n \right)+2h_n^2  &  h_n \leq t \leq 2h_n
\end{cases},
\end{equation}
where $h_n$ is the $n$th peak of $|F(t)|/\gamma g$. According to Eq. \eqref{eq: C(t) SG-TIETA}, the $n$th inter-echo interval probes $D_\mathrm{inst}(t)$ over the time interval $[0, 2h_n]$, with a peak at $h_n$. With the core experimental method described, we turn towards the problem making each echo accurate. An initial step is to isolate the direct echo pathway.

\subsection{Isolating the direct echo pathway in the time domain}

Ignoring the effects of magnetic susceptibility, surface-relaxivity \cite{Mitra1993}, and spin-spin ($T_2$) relaxation (for the meantime), the predominant source of extraneous signal behavior is off-resonance coherence transfer pathways (CTPs). When the bandwidth of Larmor precession frequencies spanned by an SG exceeds the bandwidth of $\pi$-pulses, every pulse is slice-selective and excites all CTPs. \cite{Keeler2010} For SG-CPMG measurements, the number of refocused off-resonance CTPs grows exponentially with $N$ ($\sim 3^{N}$)\cite{Hurlimann2001}, resulting in significant deviations from the expected echo attenuation. \cite{Ross1997, Goelman1995, Hurlimann2001, Song2002} Phase cycling remediation schemes require $\sim 2^N$ steps \cite{Baltisberger2012} and are thus infeasible. Unconventional approaches such as time-based avoidance of CTPs \cite{Song2005, Sigmund2007} become necessary. An SG aids these time-based approaches by acting as an always-on crusher gradient. The minimum time separation to avoid undesired signal, $\tau_{\mathrm{sep}}$, is shortened. Specifically, $\tau_{\mathrm{sep}}$ is constrained by $\tau_{\mathrm{sep}} \ge \tau_p$, where $\tau_p = 2\pi/(\gamma g\Delta z)$ is the length of the $\pi$-pulse and $\Delta z$ is the resulting slice thickness. The echo width under an SG is on the order of $\tau_p$ \cite{Casanova2011} such that this constraint can be understood as avoiding undesired echo overlap. For hard $\pi$-pulses, $\tau_p$ is on the order of $\sim\mu$s.  

In the interest of acquiring an accurate direct echo CTP, we should ask: What choice of $m_j$, $\tau$, and $\delta$ separates off-resonance CTPs from the direct echo CTP by $\ge \tau_p$? Consider that CTPs are piece-wise linear functions in $F(t)$. The three magnetization states for a spin-$1/2$ nuclei (in shorthand: $M \in \{+, -, 0\}$ \cite{Song2002}) correspond to $+\gamma g$, $-\gamma g$, and $0$ slopes, respectively (see Fig. \ref{fig: pulse seq}b). Refocusing of undesired signal occurs when the summed difference between an off-resonance $F(t)$ and the direct echo $F(t)$, $\textstyle \sum\Delta F(t)$, equals 0. Rules for $m_j$, $\tau$, and $\delta$ are developed. Singly stimulated echoes (e.g., $+0-$) arise due to two $h_n$ matching. Thus:
\begin{enumerate}[label=(\roman*)]
\item Absolute $F(t)$ heights, or $h_n$, may not be repeated. 
\end{enumerate}
Next, consider CTPs that see the initial $\pi/2$-pulse. These CTPs can alter $\sum\Delta F(t)/ \gamma g$ by one of $\left\{0, 1, 2\right\}\, \times\,  (-1)^{j-1}(2\tau +m_j\delta)$. Accounting for $\sum\Delta F(t) = 0$ with up to four non-zero terms:
\begin{enumerate}[label=(\roman*)]
    \setcounter{enumi}{1}
    \item $m_j$ and $m_{j\pm\Delta j}$ with odd $\Delta j$ may not be the same.
    \item Twice any $m_j$ may not equal the sum of $m_{j+\Delta j}$ and $m_{j-\Delta j}$ for even $\Delta j$. 
    \item Any two $m_j$ with even $j$ may not equal the sum of any two $m_j$ with odd $j$. 
    \item Twice any $m_j$ with even $j$ may not equal the sum of any two $m_j$ with odd $j$, and vice versa.
\end{enumerate}
Another class of off-resonance CTPs is associated with the introduction of transverse magnetization from spins that incorrectly see $\pi$-pulses as initial $\pi/2$-pulses. These CTPs emerge at the time of $\pi$-pulses and thus invariably start with $\sum \Delta F(t)$ containing an odd multiple of $\tau$ (i.e., $\tau + 2j\tau$). Choosing $\tau$ and $\delta$ such that $(\tau \bmod \delta) = \delta/2$ ensures that $\sum \Delta F(t) \ge\delta/2$. Incorporating $\tau_p$:
\begin{enumerate}[label=(\roman*)]
    \setcounter{enumi}{5}
    \item $\delta$ and $\tau$ satisfy $(\tau \bmod \delta) = \delta/2$ and $\delta > 2\tau_p$.
\end{enumerate}
Note that $(\tau \bmod \delta) = 0$ results in the refocused CTPs in Fig. \ref{fig: pulse seq}b. The sequence in Fig. \ref{fig: pulse seq} does, in fact, satisfy rules (i--v). This limited ruleset (i--vi) may be sufficient to ostensibly avoid off-resonance effects considering that singly stimulated echoes are known to be the most significant contributor to SG-CPMG off-resonance effects \cite{Hurlimann2001, Song2002}.

The generation of $m_j$ that satisfies rules (i--v) is discussed in the Supplementary Material (SM) Section\ I. Python code is provided. A solution for $\tau$, $\delta$, and $m_j$, used throughout, is 
\begin{equation}\label{eq: timings}
\begin{gathered}
    \tau = 49 \,\mu \mathrm{s}, \;\;\;\; \delta = 14\, \mu \mathrm{s},\\
    m_j = \left\{\begin{matrix}
    1 & 3 & 6 & 7 & 10 & 12 & 11 & 15 & 20 & 21 \\ 
    24 & 26 & 20 & 21 & 33 & 35 & 33 & 34 & 33 & \hdots
    \end{matrix}\right\}, 
\end{gathered}
\end{equation}
which gives time sensitivity over $t\sim50-500\;\mu$s,
\begin{equation}\nonumber
    h_n = \left\{\begin{matrix}
    49 & 63 & 77 & 105 & 91 & 147 & 119 & 133 \\
    175 & 203 & 189 & 245 & 217 & 161 & 231 & 329 \\ 
    259 & 301 & 273 & 287 & \hdots
    \end{matrix}\right\}\, \mu \mathrm{s}.
\end{equation}

\section{Experimental Setup}

NMR measurements were performed at $\textbf{B}_0=0.3239\ \mathrm{T}$ (proton $\omega_0=13.79\ \mathrm{MHz}$) using a PM-10 NMR MOUSE single-sided permanent magnet \cite{Eidmann1996} (Magritek, Aachen Germany) and a Kea 2 spectrometer (Magritek, Wellington, New Zealand). The decay of the magnetic field with distance from the magnet produces a strong SG of $g = 15.3\ \mathrm{T/m}$ ($650\ \mathrm{kHz/mm}$). Measurements used a home-built test chamber and a $13\times2$ mm solenoid RF coil and RF circuit. Additional information concerning the experimental setup can be found in Williamson et al. \cite{williamson2019magnetic} The SG-TIETA pulse program was written in Prospa V3.22 by modifying the standard CPMG sequence. See SM Section IV for details. 
 
For measurements on twice-distilled water, decane (Sigma-Aldrich, St. Louis, MO, USA.), 1-octanol (Sigma-Aldrich), and dodecamethylcyclohexasiloxane (D6) kinematic Viscosity = 6.6 cSt @ 25°C (Gelest, inc. Morrisville, PA, USA.), the liquids were transferred to 2 cm glass capillary sections (1.1 mm OD, Kwik-Fil\texttrademark, World Precision Instruments, Inc., Sarasota, FL, USA) and the capillaries were sealed with a hot glue gun. 
 
For measurements on yeast (\textit{S. cerevisiae}), 1.72 g of dry yeast was mixed in 10 ml of tap water and stored in a 50 ml tube with the lid screwed on loosely to allow gas to escape. After three days (72 hours), the yeast was re-suspended and samples were taken for NMR experiments and for cell density measurement. The density of the first sample (yeast \#1) was measured to be $2.84\times 10^9$ cells/ml using a hemocytometer. The remaining yeast was centrifuged, the pellet was re-diluted to 2X the initial concentration, and a second sample (yeast \#2) was taken for NMR experiments. Immediately prior to experiments, yeast was transferred to 2 cm KrosFlow\textsuperscript{\textregistered} Implant Membrane sections (500,000 Dalton molecular weight cut off, 1 mm outer diameter, SpectrumLabs, Waltham, MA, USA) and the membranes were sealed by pinching the membrane with heated forceps. Yeast was kept from drying out by performing measurements immediately upon filling and sealing the capillary and by lining the inside with wet tissue paper to increase humidity.

Experiments were performed at ambient temperature. Sample temperature was monitored with a fiber optic sensor (PicoM Opsens Solutions Inc., Qu\'ebec, Canada). The average temperatures of the samples during the course of the experiments were 25$\mathrm{^\circ C}$, 22$\mathrm{^\circ C}$, 22$\mathrm{^\circ C}$, 23$\mathrm{^\circ C}$, and 24$\mathrm{^\circ C}$ for the water, decane, octanol-1, D6, and yeast, respectively.

\section{Results and Discussion}

\begin{figure}
\includegraphics{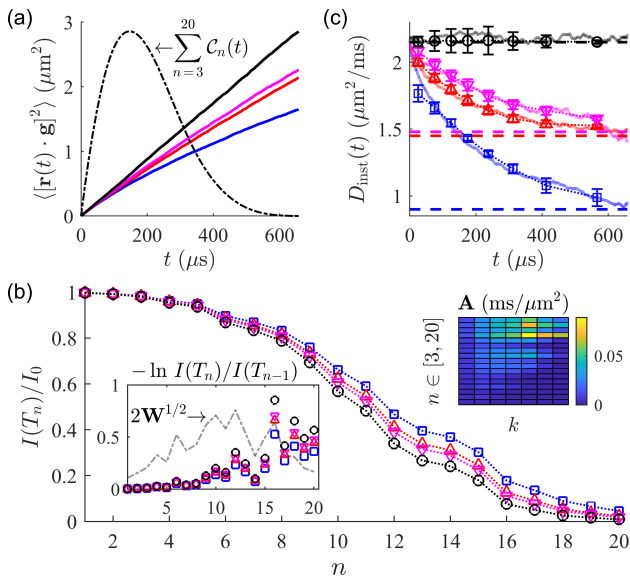}
\caption{\label{fig: Monte Carlo} LLS inversion on Monte Carlo simulated data. (a) Simulated $\langle \left[ \textbf{r}(t) \cdot \hat{\textbf{g}} \right]^2\rangle$ ($D_0 = 2.15\;\mu\mathrm{m}^2/\mathrm{ms}$) for free diffusion (black) and restricted geometries (see Fig. S1). $\sum \mathcal{C}_n(t)$ (a.u.) for Eq. \eqref{eq: timings} is overlaid, omitting the first two echoes. (b) Echo decays simulated from same color curves in (a) for $\gamma g = 4.093 \; \mu\mathrm{m}^{-1}\mathrm{ms}^{-1}$. Insets show \textbf{B} with $2\textbf{W}^{1/2}$ for the black curve and \textbf{A} for $\Delta t(k) = \{50, 50, 50, 50, 75, 75, 125, 139\}\; \mu$s. (c) $D_{\mathrm{inst}}(t)$ from the gradient of $\langle \left[ \textbf{r}(t) \cdot \hat{\textbf{g}} \right]^2\rangle$ curves in (a) (solid lines) compared to $\textbf{X}$ inverted from decays in (b) with added Gaussian noise (SNR = 25). Err. bars $= \pm 1$ SD from 100 replications. Initial \textbf{X} guesses were $D_0$ for the first point and $D_\infty$ (dashed lines, $\{0.9, 1.45, 1.48\} \;\mu\mathrm{m}^2/\mathrm{ms}$) for remaining points. $\lambda = 2\times 10^{-6}$, selected manually. See Eq. (S1) for $\boldsymbol{\Gamma}$.}
\end{figure}

As an initial proof-of-principle, the LLS inversion described in Eq. \eqref{eqn: lin prog} was performed on noisy SG-TIETA decays generated using the timings in Eq. \eqref{eq: timings} and $\langle \left[ \textbf{r}(t) \cdot \hat{\textbf{g}} \right]^2\rangle$ curves from Monte Carlo simulations \cite{Sousa2018}, shown in Fig. \ref{fig: Monte Carlo}. Further details and MATLAB code are provided in SM Section\ II. Inverted $\textbf{X}$ values are shown to be accurate and robust to noise. No systematic errors other than potential over-regularization are observed.

Echo-to-echo accuracy remains experimentally difficult to achieve, however. Consider the signal decay due to $T_2$ and pulse inaccuracy effects. \cite{Song2002} Relaxation may be ignored if $(2\tau + m_j\delta) \ll T_2 \; \forall j$. Diffusion-weighted $T_2$ values for each sample, as shown in Table \ref{tab: relaxation}, indicate that this condition holds for all pure liquids studied here. However, the yeast is described by a distribution of $T_2$ with a 5\% water population with $T_2$ similar to $\mathrm{max}\left\{2\tau + m_j\delta\right\} = 539 \;\mu \mathrm{s}$. The instantaneous diffusion measurements of yeast may be slightly weighted by $T_2$ relaxation. See SM Section IV.D for relaxation measurement methods and a $T_2$ distribution analysis for yeast \#2. Unlike $T_2$, pulse inaccuracy cannot be ignored. Inaccuracy effects may be described using an $n$-dependent pulse accuracy factor, $A_p(n)$. \cite{Song2002} The signal is then corrected as $I(T_n)/I_0 \times \left[1/\,\prod_{l=1}^n A_p(l)\;\right]$, assuming total avoidance of off-resonance CTPs via rules (i--vi).

\begin{table}[ht]
\caption{\label{tab: relaxation} Relaxation times. }
\begin{ruledtabular}
 \begin{tabular}{|l c c |} 
 Sample & $T_1$ [ms] & $T_2$ [ms] \\ 
 \hline 
 water & 3250 & $133$ \\ 
 decane & 1340 & $166$ \\ 
 octanol-1 & 440 & $217$ \\ 
 D6 & 788  & $292$ \\ 
yeast \#1 (2.84B cells/mL) & 625 & $59$ \\ 
 yeast \#2 (5.68B cells/mL) & 319 & $41$ \\ 
 \end{tabular}
\end{ruledtabular}
\end{table}

\begin{figure}
\includegraphics{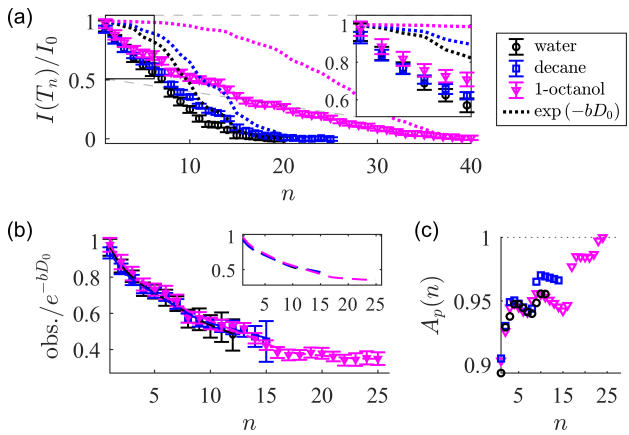}
\caption{\label{fig: pulse acc} SG-TIETA $A_p(n)$ calibration using pure liquids. (a) Observed SG-TIETA decays compared to $\exp{(-bD_0)}$. $D_0$ was measured (see SM Section IV) in legend order as $2.22 \pm 0.01, 1.27 \pm 0.01$, $0.121 \pm 0.002\; \mu\mathrm{m}^2/\mathrm{ms}$. Err. bars $=\pm1$ SD for 25, 38, 70 repetitions, respectively. (b) Decay vs. $\exp{(-bD_0)}$ ratio truncated at $n = 12, \,15, 25$. Inset shows cubic spline fits. (c) $A_p(n)$ approximated using adjacent fitted ratios.}
\end{figure}

\begin{figure}[hbt!]
\includegraphics{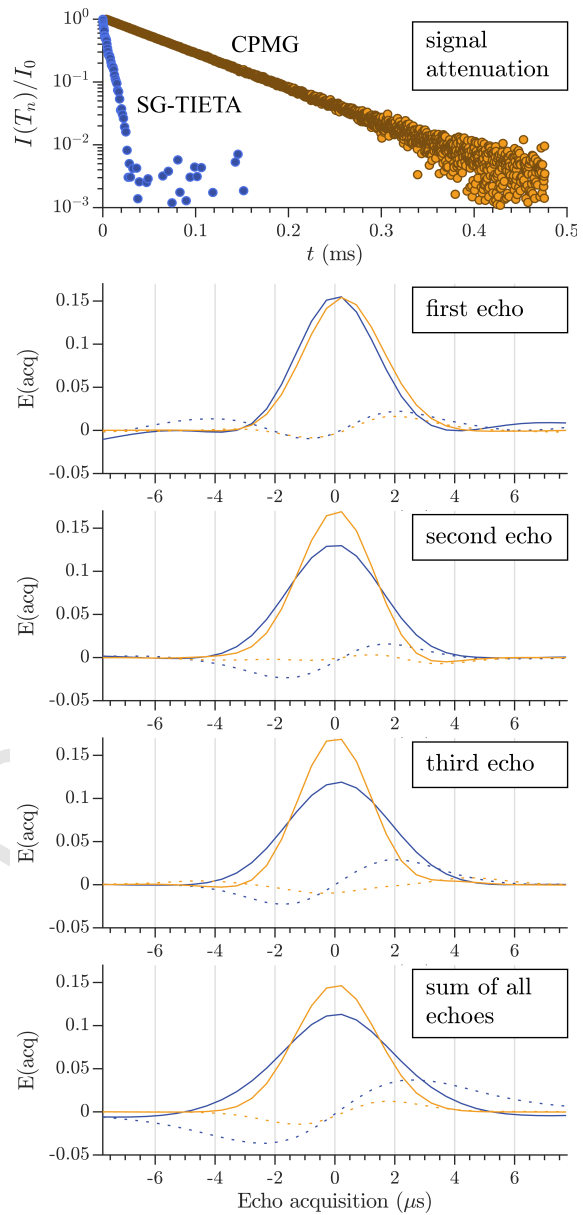}
\caption{Comparison of the direct echo SG-TIETA (blue) and CPMG (orange) attenuation and echo shape for 1-octanol with $2\tau$ = TE = 98 ms. Echo shapes show the real (solid lines) and imaginary (dotted lines) signal normalized by the area under the real signal curve in a 16 $\mu$s window. The CPMG echo width decreases with $n$ and stabilizes around $n = 3$, consistent with the approach to asymptotic behavior described by H{\"u}rlimann \& Griffin.\cite{hurlimann2000spin} In contrast, the SG-TIETA echo width \emph{increases} and stabilizes around $n = 3$, consistent with the direct echo CTP being preferential to the on-resonance signal. See Figs. S7 and S8 for all echo shapes and an exemplar echo decay, respectively.}  
\label{fig: compare CPMG}
\end{figure}

Calibration $A_p(n)$ values were approximated from decays of pure liquids, shown in Fig. \ref{fig: pulse acc}. Echo amplitudes were calculated as the sum of all real signal points within the echo window, which was set to 16 $\mu$s. All decays were normalized to the first echo and each repetition consisted of 32 summed (i.e., signal-averaged) scans. To elucidate the expected $A_p(n)$ behavior, we consider the spatial, i.e., slice effects. The bandwidth/slice excited by refocusing $\pi$-pulses has inconsistent frequency content \cite{geil1998measurement,hurlimann2000spin} such that $A_p(n) < 1$. With each pulse, spins which rotate by angles other than $\pi$ do not refocus until only a stable, central slice remains. In the time domain, this slice-thinning and loss of frequency content is expected to broaden the echo width, which we experimentally verify in Fig. \ref{fig: compare CPMG}. Based on the evolution of the echo shape, $A_p(n)$ should sharply increase then taper. This behavior is observed in Fig. \ref{fig: pulse acc}c and is consistent across liquids of vastly different $D_0$. Similar $A_p(n)$ values were also obtained for $\tau = 77 \; \mu$s (see SM Section III, Fig. S5), further supporting that $A_p(n)$ is independent of the diffusion weighting. 

Another pre-processing step is designed to mitigate the effects of early $A_p(n)$ variability and to ensure the non-negativity of \textbf{B} and \textbf{W} entries. A piece-wise linear fit of adjacent $\ln\, \bm{(}I(T_n)/I_0\bm{)}$ points vs.\ $b$ is performed, specifying (1) an intercept with $[b, \ln\, \bm{(}I(T_n)/I_0\bm{)}] = [0,0]$, (2) no slope exceeds $D_0$, and (3) the piece-wise slopes decrease monotonically (i.e., $D_\mathrm{inst}(t)$ has non-negative concavity). Altogether, SG-TIETA decays are analyzed in five steps: (1) summing $32\times$, (2) normalization to the first echo, (3) $I(T_n)/I_0 \times \left[1/\prod_{l=1}^n A_p(l)\right]$ correction, (4) a constrained log $b$ domain fit to the repetition(s), and, finally, (5) the LLS inversion. This pipeline was applied to SG-TIETA decays of D6 and yeast using the 1-octanol and water $A_p(n)$ values obtained in Fig.\ \ref{fig: pulse acc}c, respectively. 

\begin{figure}
\includegraphics{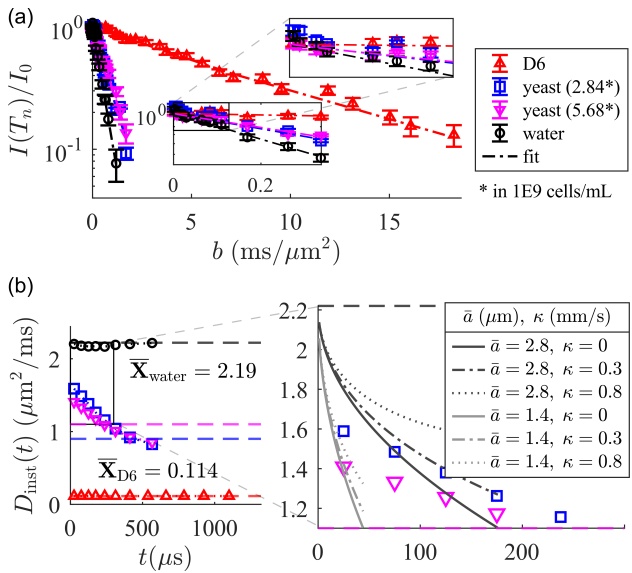}
\caption{\label{fig: key results} SG-TIETA decays and inverted \textbf{X} for D6, yeast, and water. (a) Decays analyzed as described in the text. See SM Section\ II for fitting procedures. Err. bars = $\pm$ 1 SD for, in legend order, 38, 3, 4, 25 repetitions truncated at $N = 34, 17, 17, 15$, respectively. Note erratic early $A_p(n)$ behavior. (b) $\textbf{X}$ solutions. Inversion parameters were identical to Fig. \ref{fig: Monte Carlo} other than $\Delta t(k)$ for D6. Initial guesses of $D_0 = 2.22$, $D_\infty = \{0.9, 1.1\}\; \mu\mathrm{m}^2/\mathrm{ms}$ for yeast and $D_0 = D_\infty = 0.114 \; \mu\mathrm{m}^2/\mathrm{ms}$ for D6 were provided. Zoomed plot compares short-time $D_{\mathrm{inst}}(t)$ plotted up to $t < 0.08\times\tau_D$.}
\end{figure}

Results are summarized in Fig.\ \ref{fig: key results}. The $D_{\mathrm{inst}}(t)$ for  D6 -- with $D_0 = 0.114 \pm 0.008\; \mu\mathrm{m}^2/\mathrm{ms}$ -- is expectedly flat, thus validating the $A_p(n)$ correction. The $D_{\mathrm{inst}}(t)$ for yeast are the key results of this Communication. For comparison, the short-time $D_\mathrm{inst}(t)$ predicted by Eq.\ \eqref{eq: Sen permeability short-time} (i.e., $d\left[tD(t)\right]/dt$) is plotted for several $S/V$ and $\kappa$ values. Fig. \ref{fig: key results}b indicates that $\bar{a} \simeq 2 \; \mu$m and that a doubling of the cell density approximately halves $\bar{a}$. For yeast's $\approx 4 \;\mu$m spherical diameter \cite{Suh2003}, $\bar{a}$ is calculated as 42 and 21 $\mu$m at these cell densities, suggesting contributions from \emph{sub}-cellular length scales. This ensemble $\bar{a}$ estimate of $\simeq 2 \; \mu$m ($\equiv V/S \simeq 3 \;\mu$m) is within the range of estimates ($\bar{a} \sim 1 - 5 \;\mu$m) reported in previous PFG and SG DW-NMR studies of similar yeast densities. \cite{Suh2003, Tanner1968, Aslund2009, Mazur2020, Karunanithy2019} 

Several factors may contribute to the discrepancy between the experimental and the predicted short-time $D_\mathrm{inst}(t)$. On the numerical side, over-regularization may artificially flatten $D_\mathrm{inst}(t)$ at short times, as shown in Fig. \ref{fig: Monte Carlo}. Values of \textbf{X} are also plotted inexactly at the midpoints of $\sum_k \Delta t(k)$. On the theoretical side, Eq. \eqref{eq: Sen permeability short-time} does not include a term for the curvature, which may be significant at these timescales for sub-cellular water. Consider, also, the confounding effects of $T_2$, surface relaxation, the Gaussian phase approximation, and the assumption of echo number translation invariance (i.e., that each $C_n(t)$ is presumed to start from $t = 0$). Echo number translation invariance does not hold for spatially heterogeneous microenvironments. If different water pools exhibit varying decay rates, the relative signal contributions will depend on the echo number, $n$. Indeed, spatial heterogeneity and the resulting weighting towards slowly decaying water pools at larger $n$ may explain the convergence of the yeast \textbf{X} values. These yeast results are thus non-quantitative. Nonetheless, the sensitivity of SG-TIETA to \emph{apparent} microstructural features is clear.

\section{Conclusions}

We have developed a real-time protocol to measure time-varying diffusion. Inversion for $D_{\mathrm{inst}}(t)$ from experimental SG-TIETA decays is demonstrated. An approximately 1-minute ($32\times$ repetition time of $2$s) experiment is described. In contrast to conventional temporal diffusion spectroscopy methods, the single-shot nature of SG-TIETA permits true signal averaging in order to improve SNR. A \textit{post hoc} $A_p(n)$ correction is proposed to improve the quantitative accuracy of the method. To support the validity of this correction, we present preliminary evidence in the observed echo shape behavior and in the consistency of $A_p(n)$ values across different diffusion weightings. Regarding potential applications, SG-TIETA at this $g$ can probe porous media microstructure on micron length scales, i.e., over sub-millisecond timescales. SG-TIETA can also be used to study phenomena associated with other long-range correlations, e.g., polymer dynamics \cite{callaghan1992evidence}, the glass transition \cite{williamson2019glass}, and high P\`{e}clet fluxes driven by flagella \cite{Short2006}, which likewise exhibit time-dependence in this sub-millisecond range. The methods contained in this Communication may open new avenues of research within DW-NMR. 

\section*{Supplementary Material}

In the supplementary material, we include sections containing (I) Python code to generate $m_j$, (II) representative MATLAB code for the Monte Carlo simulations, fitting procedures, and LLS inversion, (III) a replication of the analysis in Figs. \ref{fig: pulse acc} and \ref{fig: key results} using SG-TIETA decays for $\tau = 77\;\mu$s, and (IV) additional NMR experimental methodology, which includes all echo shapes and a stitched echo decay for 1-octanol.

\begin{acknowledgements}
The authors would like to thank Dr. Dan Benjamini and Dr. Michal Komlosh for helpful discussions concerning numerical programming and time-based avoidance of unwanted coherence pathways, respectively. 
 
TXC, VW, RR, and PJB were supported by the IRP of the NICHD, NIH. TXC is a graduate student in the NIH-Oxford-Cambridge Scholars Program. NHW was funded by the NIGMS PRAT Fellowship Award \#FI2GM133445-01. PJB, TXC, NHW, and VW conceptualized the work; TXC developed the theory and performed computations; NHW, VW and RR designed and performed experiments; NHW and TXC analyzed data; TXC prepared the manuscript; PJB supervised the project. All authors edited the manuscript.
\end{acknowledgements}

\section*{Data Availability}
The data that support the findings of this study are available
from the corresponding authors upon reasonable request. The SG-TIETA pulse program and macro can be downloaded from the GitHub repository: \url{https://github.com/nathanwilliamson/SG-TIETA}.

\end{document}